\documentclass[english,showpacs,twocolumn,aps,reprint]{revtex4-1}
\usepackage{amsmath}
\usepackage{mathtools}
\usepackage{amssymb}
\usepackage{color}
\usepackage{graphics}
\usepackage{epsfig}
\usepackage[normalem]{ulem} 
\usepackage{cancel}
\usepackage[mathlines]{lineno}
\graphicspath{{Figures/}}
\usepackage{svg}
\usepackage{xtab,afterpage,longtable}
\usepackage{booktabs}  
\usepackage{ltablex}
\usepackage{enumitem}
\relax
\usepackage{float}
\usepackage{orcidlink}
\usepackage{hyperref}
\hypersetup{colorlinks=true,linkcolor=blue,urlcolor=blue,citecolor=blue}

\begin{document}
\title{Improved Fluid Modeling of Space Debris Generated Ion-Acoustic Precursor Solitons}
\author{Ajaz Mir\orcidlink{0000-0001-5540-4967}}
\email{ajazmir.physics@gmail.com} 
\affiliation{Department of Physics, Islamic University of Science \& Technology, Awantipora 192122, Jammu \& Kashmir, India} 
\author{Abhijit Sen\orcidlink{0000-0001-9878-4330}}
\affiliation{Institute for Plasma Research, Gandhinagar 382428, Gujarat, India} 
\affiliation{Homi Bhabha National Institute, Anushaktinagar, Mumbai 400094, India}
\author{Pintu Bandyopadhyay\orcidlink{0000-0002-1857-8711}}
\affiliation{Institute for Plasma Research, Gandhinagar 382428, Gujarat, India} 
\affiliation{Homi Bhabha National Institute, Anushaktinagar, Mumbai 400094, India}
\author{Sanat Tiwari\orcidlink{0000-0002-6346-9952}}
\affiliation{Department of Physics, Indian Institute of Technology Jammu, Jammu 181221, Jammu \& Kashmir, India}
\author{Chris Crabtree\orcidlink{0000-0002-6682-9992}}
\affiliation{Naval Research Laboratory, Washington, DC 20375, USA}
\author{Gurudas Ganguli\orcidlink{0000-0002-7733-6683}}
\affiliation{Naval Research Laboratory, Washington, DC 20375, USA}
\date{\today}
\begin{abstract}
This study reexamines the excitation of ion-acoustic precursor solitons by a supersonically moving charged debris object, incorporating two previously overlooked physical factors: the dynamic charging of the debris and the impermeable nature of its surface. The influence of charging dynamics is explored using an enhanced one-dimensional fluid-Poisson model, where the source charge is treated as a dynamical variable and solved self-consistently alongside the core plasma equations. By comparing these results with prior fixed-charge models, we evaluate the effects on soliton onset and propagation, finding that charging dynamics does not hinder soliton generation or evolution. To assess the impact of the impermeability of debris surface, a two-dimensional fluid model simulates the interaction between an electrostatically biased, impenetrable object and a flowing plasma. Modeling the object as an infinite wall disconnects the upstream and downstream plasma regions, forming a sheath without solitons---consistent with earlier fluid and particle-in-cell  simulations. However, replacing the wall with a finite object enables plasma flow around it, restoring upstream-downstream  connectivity and naturally generating precursor solitons.
\end{abstract}
\maketitle
\section{Introduction}
\label{Intro}
Precursor nonlinear structures in the form of solitons excited by a charged object traveling supersonically through a plasma have received significant research attention of late because of their potential as an indirect means of detecting and tracking small-sized debris objects~\cite{Bernhardt_POP_2023, Truitt_AMOS_2023}. 
The basic concept of such nonlinear ion-acoustic excitations in a plasma medium was demonstrated theoretically in a simple one-dimensional (1D) fluid-Poisson model calculation leading to a paradigmatic nonlinear partial differential equation, the forced Korteweg-de Vries (fKdV) equation~\cite{Sen_ASR_2015}. The fKdV model yielded a variety of driven nonlinear solutions in the form of precursor solitons, “pinned” solitons, wakes and dispersive shocks. The scale sizes and amplitudes of these nonlinear structures for ionospheric conditions are within the detection capabilities of ground-based radars or in situ sensors~\cite{Zabotin_RS_2001,Rietveld_PS_2008,Sinha_JGRSP_2010}. Hence, they can provide an indirect means of detecting debris objects. 
Subsequent studies have established the existence of such solitons in a variety of theoretical models~\cite{Sanat_POP_2016A, Truitt_JSR_2020A, Truitt_JSR_2021, Ajaz_PRE_2025} as well as in molecular dynamics (MD)~\cite{Sanat_POP_2016B} and Particle-In-Cell (PIC)~\cite{Atul_NJP_2020, Vikram_PRE_2023, Atul_PPCF_2019} simulations. Their existence has also been demonstrated in controlled laboratory experiments~\cite{Surabhi_PRE_2016,Garima_POP_2019,Krishan_POP_2024}. In most of the theoretical studies, the debris object has been modeled as a Gaussian function of space that carries a constant charge determined by the ambient plasma conditions.  Neglecting the initial charging dynamics of the source has been physically justified on the grounds that the charging process takes place on the ion time scale, $\omega_{pi}^{-1}$ (where $\omega_{pi}$ is the ion plasma frequency), which is much faster than the ion-acoustic time scale on which the solitons develop.  In other words, much before the initiation of soliton formation and subsequent evolution, the source would have acquired its equilibrium charge value. The Gaussian form of the debris charge density mimics the Debye shielded nature of a charged object in a plasma and has been successfully used in fluid simulations as well as in model nonlinear equations such as the fKdV equation~\cite{Sen_ASR_2015, Hartzell_PRE_2025} or the forced Kadomtsev–Petviashvili (fKP) equation to study the excitation and propagation of precursor solitons~\cite{Ajaz_PRE_2025}. 
\paragraph*{}
There has been some recent criticism, that are based on findings from PIC simulations~\cite{Lira_JSR_2024}, of both the simplifying assumptions, namely, the constant charge on debris and the Gaussian form of the debris charge density. It has been suggested that the charging process could lead to large scale undamped potential fluctuations that might impede the generation of solitons.  It has also been pointed out that the Gaussian  source term is transparent to the plasma allowing it to flow freely through the debris object whereas real life debris objects, composed of solid material, would reflect the plasma back and thereby disconnect the regions of the plasma in the front and rear regions of the debris. Simulating such a situation by using an infinite wall separating the two regions it was shown, in PIC simulations~\cite{Lira_JSR_2024}, that one would not have any solitons but only the formation of a plasma sheath on the wall. In view of these reservations we have carried out an improved fluid simulation study where both these factors are taken into account while investigating the excitation of precursor solitons. The charging dynamics is incorporated in an extended 1D fluid model in a self-consistent manner by solving the source charging equation simultaneously with the plasma dynamical equations. A comparison of the results with those obtained in the past for a fixed source charge show that the charging dynamics does not have any significant inhibitory impact on the generation of precursor solitons. An investigation is also made of the early time density and potential fluctuation dynamics of the two cases to provide a deeper insight into the soliton formation process as well as to carry out a critical comparison with the recent contrarian PIC simulation findings~\cite{Lira_JSR_2024}.
\paragraph*{}
To study the impact of the impermeable nature of the plasma facing surface of the debris we carry out a two dimensional (2D) simulation where a finite sized object, a line source, maintained at a constant potential is made to interact with a flowing plasma. Appropriate boundary conditions are used to mimic the impermeable nature of the source by making the plasma reflect off the surface. When the line source is extended to represent an infinite wall we get a sheath formation in agreement with the PIC results. However, for a finite source that permits the plasma to flow around it, as would be the case for a realistic debris object speeding through an ambient space plasma, we find that solitons are generated and the impermeable nature of the plasma facing surface of the debris does not inhibit their formation. Our results thus provide a sound physical basis for the simplifying assumptions of the earlier theoretical models~\cite{Sen_ASR_2015, Sanat_POP_2016A, Ajaz_PRE_2025}.
\paragraph*{}
The paper is organized as follows. 
In Section~\ref{Charging_model} we describe the self-consistent model equations that incorporate the dynamical charging of the debris object and present the simulation results showing the creation and evolution of the precursors in the presence of charging. We also show a comparison with previous results obtained for a source with a time independent fixed charge. Section~\ref{Solid_Wall} is devoted to the investigation of the impact of an impermeable plasma facing surface on the generation of solitons that includes a discussion of an improved two dimensional fluid model as well as a comparison with an infinite line source and a finite line source. A short summary of our main findings with some concluding remarks are provided in Section~\ref{conclusions}.
\section{Effect of Dynamic Charging of Debris on Soliton Formation}
\label{Charging_model}
\subsection{Model Equations for One-Dimensional Study of Precursor Solitons}
\label{sec:level2}
We consider a one-dimensional fluid model to describe the dynamics of nonlinear ion-acoustic waves that are continuously generated by a supersonically moving charged source, representing a charged debris object. The basic dynamics of the system are then described by the following set of equations~\cite{Sen_ASR_2015}, 
\begin{equation}\label{CE_eqn1}
\frac{\partial n_i}{\partial t} 
+
\frac{\partial (n_i u)}{\partial x} 
= 0
\end{equation}
\begin{equation}\label{ME_eqn2}
 \left( \frac{\partial }{\partial t} 
+
u  \frac{\partial}{\partial x}  \right)u
=  -\frac{e}{m_i} \frac{\partial \phi}{\partial x} - \frac{k_B T_i}{m_i n_i}\frac{\partial n_i}{\partial x}
\end{equation}
\begin{equation}\label{PE_eqn3}
    \frac{\partial^2 \phi}{\partial x^2}   = 4\pi( n_e e - n_i e + S(x-v_d t) )
\end{equation}
\noindent
where Eq.~\eqref{CE_eqn1} is the ion continuity equation, Eq.~\eqref{ME_eqn2} is the ion momentum equation and Eq.~\eqref{PE_eqn3} is the Poisson equation. The field quantities $n_e$,~$n_i$,~$u$, and $\phi$ represent the electron density, ion density, ion velocity,  and electrostatic potential, respectively. $T_i$ is the ion temperature. 
\paragraph*{}
The electrons are assumed to have a Boltzmann distribution $n_{e} = n_{e0} \exp( e\phi / k_B T_{e})$, where $n_{e0}$ is the equilibrium electron density, $T_e$ is the electron temperature, $e$ is the electron charge and $k_B$ is the Boltzmann constant. The quantity $S(x-v_dt)$ is the charge density source representing the charged debris object moving with speed $v_d$ in the background electron-ion plasma. We take the following model form for the charged debris object,
\begin{equation}
\label{DYN_source}
   S(x-v_d t) = \frac{Q(t)}{w \sqrt{\pi}} \exp\left[ -\left( \frac{ x - x_0- v_d t}{w} \right)^2 \right],
\end{equation} 
where $Q(t)$ is the total charge on the debris object and $w$ is the width of the Gaussian profile and $x_0$ is the initial location of the source.
\paragraph*{}
Unlike past model calculations~\cite{Sanat_POP_2016A}, where the charge $Q$ was taken to be a constant quantity, in the present work we will treat $Q$ as a time-varying dynamical quantity to account for the charging dynamics of the debris in ambient plasma. The debris objects get charged due to the flow of electron and ion currents on them from the ambient plasma, which is a time-dependent process. The dynamics of charging can be modeled by the following charging equation~\cite{Jana_PRE_1993}, 
\begin{eqnarray}\label{CHG_eqn4}
\frac{d Q}{dt} = I_e + I_i  
\end{eqnarray}
where $I_e$ and $I_i$ represent the electron and ion currents. There can be additional charging processes arising from photoelectric effects or due to secondary electron emissions, \textit{etc}, which we will ignore in our present calculations. The electron and ion currents are determined from the Orbital Motion Limited (OML) theory and, for a Maxwellian plasma environment, are given by~\cite{Jana_PRE_1993},
\begin{equation}\label{EC_eqn}
    I_e 
    = - \sqrt{8} e w v_{the} n_e  \exp\left[ \frac{e}{k_B T_e} (\phi_d - \phi) \right], 
\end{equation}
\begin{equation}\label{EI_eqn}
 I_i = \sqrt{8} w e v_{thi} n_i \left[ 1 - \frac{e}{k_B T_i} (\phi_d - \phi) \right].
\end{equation}
where $\phi_d$ is the potential on the debris surface. For the moving debris source with $v_d$ much greater than the ion thermal speed, the ion current on the debris surface is given by
\begin{eqnarray}\label{EIVd_eqn}
    I_i = w e v_d n_i \left[1 - \frac{2 e}{m_i v_d^2} (\phi_d -\phi)\right].
\end{eqnarray}
The expressions for the electron and ion currents are taken from Jana {\it et al.}~\cite{Jana_PRE_1993} and further modified to reflect the 1D nature of the source. This is done by replacing $a^2$ by $w$.
We have adopted the following normalizations:
\begin{align*}
& x \rightarrow \lambda_{De} = \left( \frac{k_B T_e}{4\pi n_{e0} e^2} \right)^{1/2}; 
\hspace{0.35cm}
t \rightarrow \omega_{pi}^{-1} = \left( \frac{m_i}{4\pi n_{i0} e^2} \right)^{1/2}; \\
& (u,\, v_d) \rightarrow c_s = \left( \frac{k_B T_e}{m_i} \right)^{1/2}; 
\hspace{0.35cm}
n \rightarrow n_0 = n_{e0} = n_{i0}; \\
& (\phi,\, \phi_d) \rightarrow \phi_0 = \frac{k_B T_e}{e};
\hspace{0.5cm}
Q \rightarrow Q_0 = e n_0 \lambda_{De}.
\end{align*}
Here $\lambda_{De}$ is the plasma Debye length and $c_s$ is the ion-acoustic phase speed.
\noindent
We further adopt the ansatz $\phi_d = Q / w$ (all normalized quantities) in analogy with the potential on the surface of a sphere of radius $w$, that has charge $Q$. \\
\paragraph*{}
We then obtain the following coupled set of normalized equations for the variables $n, u, \phi, \phi_d$ that describe the dynamics of ion-acoustic waves in the presence of a source $S(x-v_d t)$ whose charge $Q(t)$ is being self-consistently evolved. \\
\begin{equation}\label{NCE_eqn1}
\frac{\partial n}{\partial t} 
+\
\frac{\partial (n u)}{\partial x}
= 0
\end{equation}
\begin{equation}\label{NME_eqn2}
   \frac{\partial u}{\partial t} 
   +\ 
   u \frac{\partial u}{\partial x} 
   =\ 
 -\frac{\partial \phi}{\partial x}  - \sigma \frac{\partial ln(n)}{\partial x}
\end{equation}
\begin{equation}\label{NPE_eqn3}
    \frac{\partial^2 \phi}{\partial x^2}  
    =  e^{\phi} - n + S(x-v_d t)
\end{equation}
The normalized form of the dynamically evolving charged source density is given by
\begin{equation}
\label{NDYN_source}
   S(x-v_d t) = \frac{\phi_d(t)}{ \sqrt{\pi}} \exp\left[ -\left( \frac{ x - x_0 - v_d t}{w} \right)^2 \right]
\end{equation} 
\paragraph*{}
The temporal evolution of the floating potential $\phi_d (t)$ for a stationary charged debris, \textit{i.e.,} $v_d=0$, is given by
\begin{equation}
\label{NCHGVd0_eqn4}
\frac{d \phi_d}{dt} = -\sqrt{8\mu} w e^{\phi_d} + \sqrt{8 \sigma} w n[1 - \frac{1}{\sigma} (\phi_d - \phi)]           
\end{equation}
where $\sigma = T_i/T_e$ and $\mu = m_i /m_e$. As can be seen, the charging rate is influenced by both the ambient plasma properties, represented by $\sigma$, $\mu$ and $n$, as well as by the debris size represented by $w$.
\paragraph*{}
For a streaming charged debris object, the debris speed $v_d$ also plays a role, and the floating potential evolution is given by,
\begin{equation}
\label{NCHGVd_eqn4}
\frac{d \phi_d}{dt} = -\sqrt{8 \mu}w e^{\phi_d} + v_d w n \left[1 - \frac{2} { v_d^2} (\phi_d -\phi)\right]
\end{equation}
In the next subsection, we will present numerical solutions of the above final set of equations consisting of \eqref{NCHGVd0_eqn4} for $v_d=0$ or \eqref{NCHGVd_eqn4} for finite $v_d$.
\paragraph*{}
Before that we obtain an approximate estimate of the equilibrium floating potential $\phi_{d0}$ and the charging time by making a linear expansion of the form $\phi_d = \phi_{d0} + \tilde{\phi}_d$, $n=1+\tilde{n}$, $\phi = \tilde{\phi}$ in equations \eqref{NCHGVd0_eqn4} and \eqref{NCHGVd_eqn4}. Here the quantities with a tilde represent small fluctuations around the equilibrium quantities~\cite{Cui_IEEE_1994}.
\paragraph*{}
Therefore, for $v_d=0$, we get,
\begin{equation}
\label{eqlbmpot1}
    \sqrt{\mu}e^{\phi_{d0}} = \sqrt{\sigma}(1 - \frac{1}{\sigma}\phi_{d0})
\end{equation}
\begin{equation}
\label{pertpot1}
    \frac{d \tilde{\phi}_d}{dt}+ w(\sqrt{8\mu}e^{\phi_{d0}}+\sqrt{8/\sigma})\tilde{\phi}_d = w\sqrt{8\sigma}[\tilde{\phi}+\tilde{n}(1 - \frac{1}{\sigma}\phi_{d0})]
\end{equation}
Taking $\sigma = 0.1$, $\mu = 1836$, $w=0.5$ and numerically solving (\ref{eqlbmpot1}) we get $\phi_{d0} = -1.9089$. From (\ref{pertpot1}), assuming $\phi_d \sim e^{-t/\tau_c}$ we can estimate the time scale $\tau_c$ on which equilibration is reached and hence the charging time to be
$$\tau_c \approx 1 /(w\sqrt{8\mu}e^{\phi_{d0}}+w\sqrt{8/\sigma}) \approx 0.37 $$.
Similarly, for $v_d \neq 0$, we get,
\begin{equation}
\label{eqlbmpot2}
 \sqrt{8\mu}e^{\phi_{d0}} = v_d  \left[1 - \frac{2} { v_d^2} \phi_{d0}\right]   
\end{equation}
\begin{equation}
 \label{pertpot2}   
 \frac{d\tilde{\phi}_d}{dt} + w(\sqrt{8\mu}e^{\phi_{d0}}+\frac{2}{v_d})\tilde{\phi}_d = w\tilde{n}v_d (1- \frac{2 \phi_{d0}}{ v_d^2})+ w\frac{2}{ v_d}\tilde{\phi}
\end{equation}
Taking $\sigma = 0.1$, $\mu = 1836$, $w=0.5$, $v_d=1.15$ and numerically solving (\ref{eqlbmpot2}) we get $\phi_{d0} = -2.9578$. From (\ref{pertpot2}) we can estimate,  
$$\tau_c \approx 1 /(w\sqrt{8\mu}e^{\phi_{d0}}+w(2 / v_d)) \approx 1.441 $$.
In both cases the charging time is of the order of the ion time scale $\omega_{pi}^{-1}$.
\subsection{Numerical Simulations of One-Dimensional Precursor Solitons}
\label{sec:level3}
The model equations have been numerically solved in the following manner. The ion continuity and momentum equations are solved using a Flux-Corrected Transport (FCT) algorithm~\cite{Boris_LCPFCT_1973}. This approach is selected for its effectiveness in resolving sharp gradients and strongly nonlinear features, that are characteristic of solitons, while suppressing numerical diffusion and nonphysical oscillations~\cite{Sanat_POP_2016A, Sanat_NJP_2012}. The self-consistent electrostatic potential is obtained by solving  Poisson's equation at each time step using a pseudo-spectral method that ensures high spatial accuracy. The charging (floating potential) equation \eqref{NCHGVd0_eqn4} or \eqref{NCHGVd_eqn4} is solved by the fourth-order Runge–Kutta (RK4) method. Our numerical code was validated by matching the linear solutions against an analytic ion-acoustic dispersion relation and also by reproducing previous nonlinear solutions~\cite{Sanat_POP_2016A} that were obtained in the absence of an explicit charging mechanism. As mentioned earlier, only the ion dynamics has been evolved explicitly, and the electrons are assumed to follow the Boltzmann distribution. The code simultaneously solves~\eqref{NCE_eqn1} - \eqref{NPE_eqn3} and \eqref{NCHGVd0_eqn4} or \eqref{NCHGVd_eqn4} for the cases $v_d = 0$ or $v_d \neq 0 $,  respectively. We have employed periodic boundary conditions for the presented simulations.
The system-size is sufficiently large $-400 \leq L_x \leq +400$ to avoid any boundary effects over domains of interest, i.e., the source and evolving precursors and wakes. The ion dynamics is resolved at 1000 part of the ion plasma period, which conveniently shows the initial transient charging of object. The inset in Fig.~\ref{Fig_1} shows the surface potential achieving a steady state within a couple of ion plasma periods. The chosen simulation timestep is also aligned with the choice of the ion-to-electron mass ratio $\mu = 1836$ for a typical electron-proton plasma.
\paragraph*{}
We now present results from numerical solutions of the model equations for various scenarios. We begin by looking at the charging dynamics and the temporal and spatial variations of the plasma density and potential when the debris source is stationary. 
\subsubsection{Stationary source, $v_d =0$}
\label{sec:level3.1}
In this case, the temporal behavior of the floating potential $\phi_d(t)$ is obtained by solving equations~(\eqref{NCE_eqn1} - \eqref{NPE_eqn3}) along with Eq.~\eqref{NCHGVd0_eqn4}, in which the dynamics of the surface potential, $\phi_d(t)$ is governed by \eqref{NCHGVd0_eqn4} (\textit{i.e.,} for a stationary source $v_d = 0$).
\begin{figure}[h!]
     \includegraphics[width=\columnwidth]{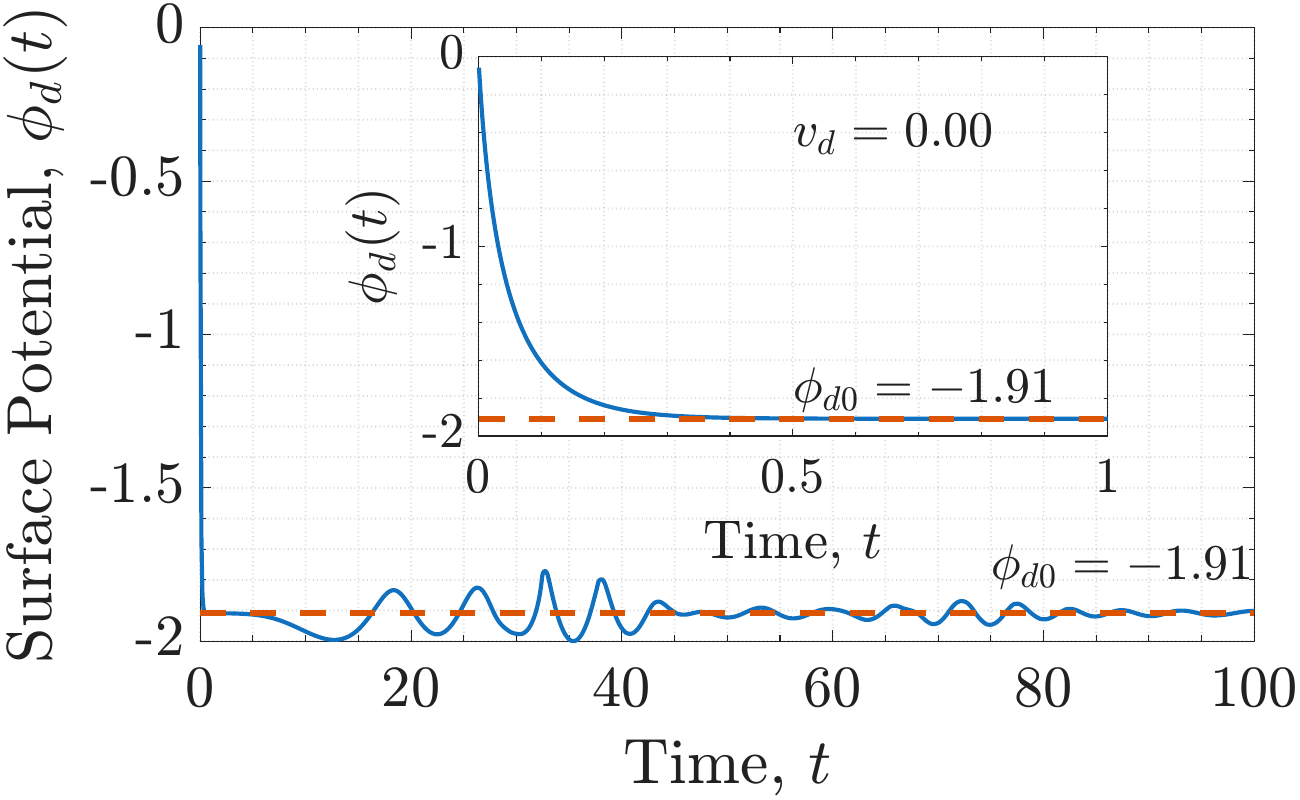}
     \caption{
     Development of floating potential on a stationary source with $\sigma = 0.1$. The inset shows the early evolution of the surface potential on the charged debris source.
     }
     \label{Fig_1}
 \end{figure}
 \begin{figure}[h!]
     \includegraphics[width = \columnwidth]{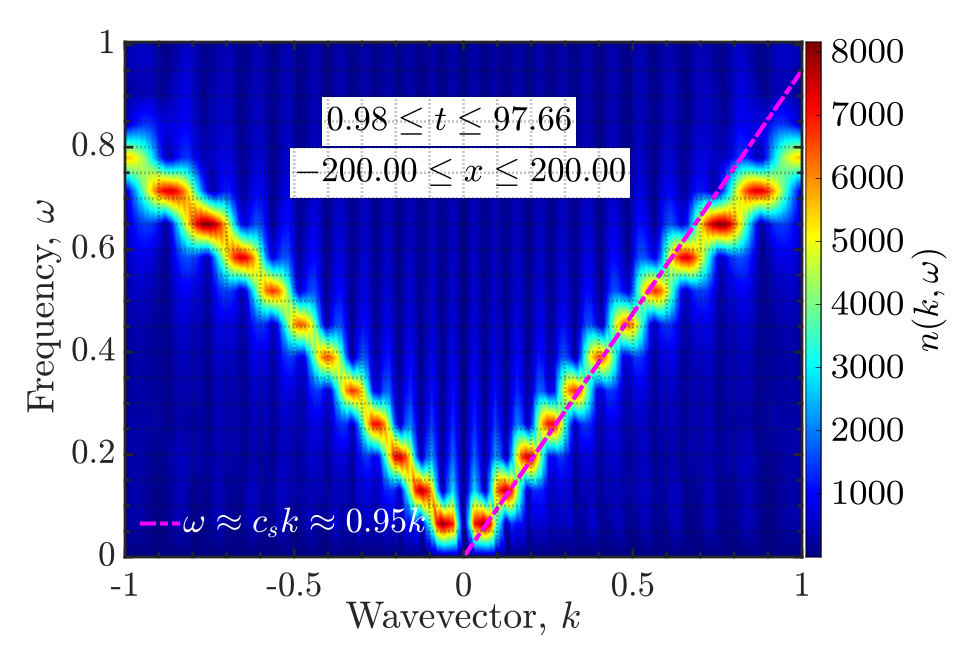}
     \caption{The $\omega-k$ dispersion relation obtained from the fluctuations in space-time during early times $t = 1.0 - 98 $ before precursors and wakes start to form. 
     }
     \label{Fig_2}
 \end{figure}
Figure~\ref{Fig_1} shows the time evolution of the surface potential (floating potential) on the debris. As can be seen, the surface potential quickly acquires an equilibrium value $\phi_{d0} = -1.91$ in agreement with the approximate value derived in the previous section. The e-folding time of this decay is about $0.40$ which is again close to the approximate analytic value. The inset in Fig.~\ref{Fig_1} is a magnified view of the early time dynamics of the charging process. In Fig.~\ref{Fig_1}, we also notice that $\phi_d(t)$ has small fluctuations in time around the equilibrium value $\phi_{d0}$. This can be understood from Eq.~\eqref{NCHGVd0_eqn4} where the RHS contains terms proportional to $\tilde{n}$ and $\tilde{\phi}$, which are fluctuating quantities and are driving the evolution of the equation. The fluctuations in $\phi_d$ are the low-frequency fluctuations of $\tilde{n}$ and $\tilde{\phi}$ composed of thermal fluctuations as well as low-frequency waves. We have identified these modes through the construction of a $\omega-k$ dispersion diagram by Fourier transforming $\tilde{\phi}$ or $\tilde{n}$ in time and space in the time interval $t = 1.0 - 98 $ during evolution.
The time-series acquired is at a location close to the debris (at $x = 0.05$) after sufficient evolution of the system such that the charging of debris is stabilized and the fluctuations start forming. The results shown in Fig.~\ref {Fig_2} indicate that these fluctuations are low-amplitude ion-acoustic waves traveling at the ion-acoustic speed. On a longer time scale, these fluctuations are amplified by a moving source, for the generation of wakes and precursor solitons ~\cite{Guru_POP_2025}. 
\subsubsection{Moving source, $v_d \neq 0$}
\label{sec:level3.2}
\begin{figure}[h!]
    \includegraphics[width= \columnwidth]{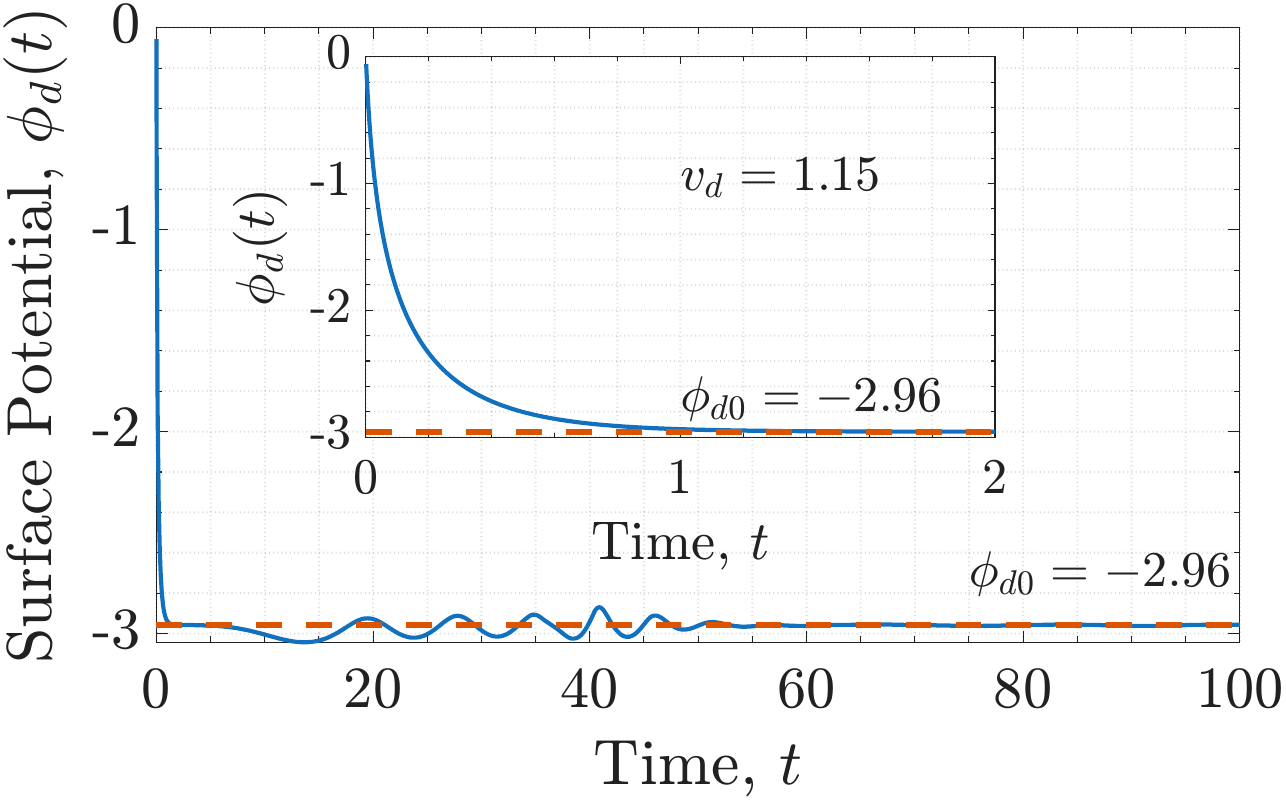}
     \caption{Temporal evolution of the floating potential on a moving source with $\sigma = 0.1$. The inset shows a zoomed view of the early time dynamics of the surface potential of the debris.}
     \label{Fig_3}
 \end{figure}

\begin{figure}[h!]
     \includegraphics[width= \columnwidth]{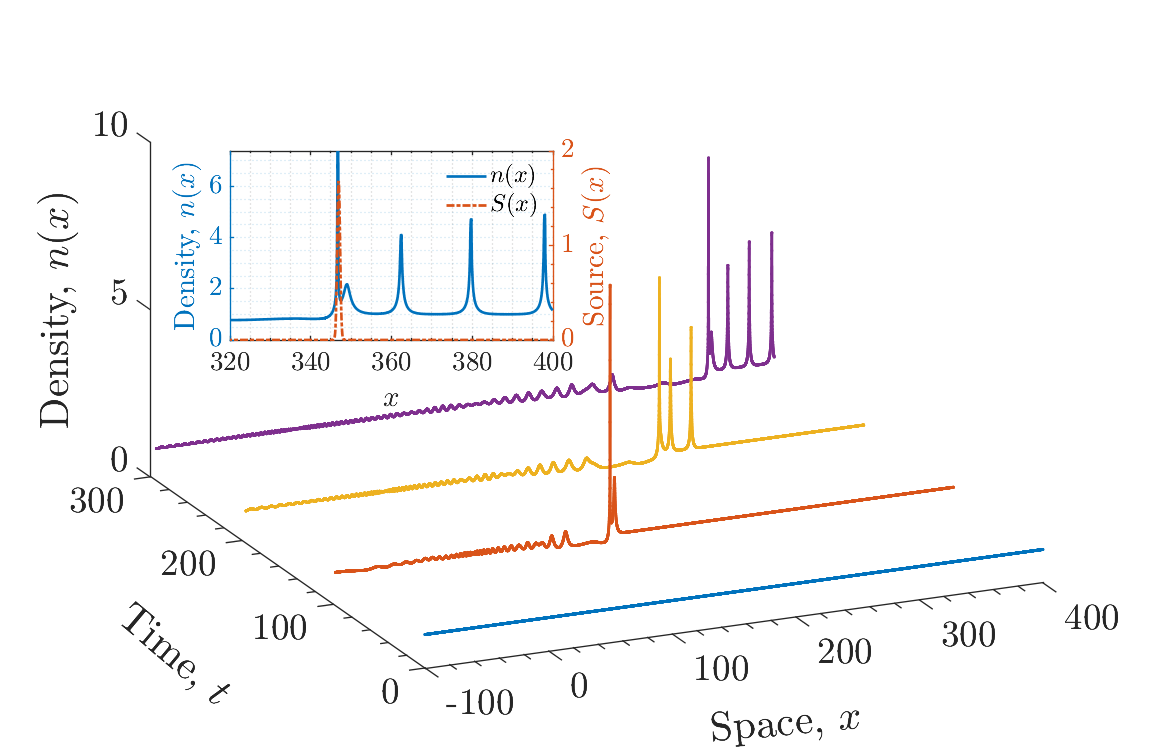}
     \caption{Spatio-temporal evolution of density {using \eqref{NCE_eqn1}-\eqref{NPE_eqn3}, and~\eqref{NCHGVd_eqn4}} showing precursors and wakes with dynamic charging of the moving source. Here $v_{d} = 1.15$, $x_0 = 10$, $w = 0.50$ and $\sigma=0$.}
     \label{Fig_4}
 \end{figure}

\begin{figure}[h!]
     \includegraphics[width= \columnwidth]{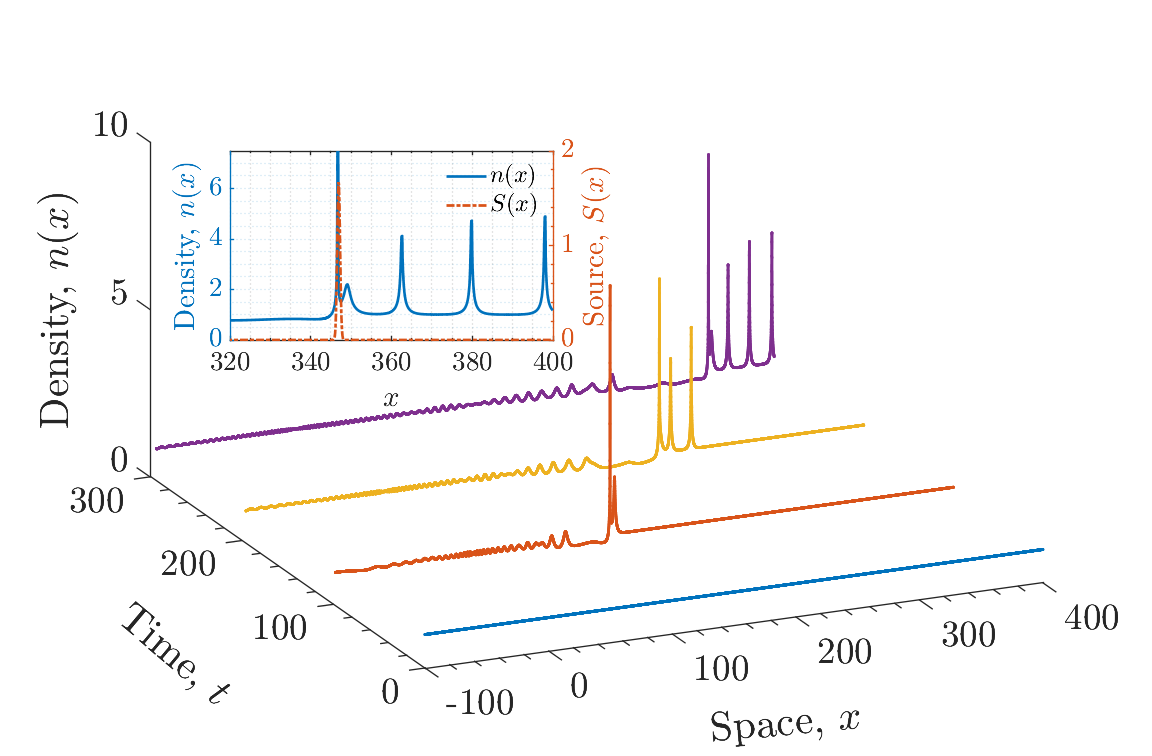}
     \caption{Spatio-temporal evolution of density {using \eqref{NCE_eqn1}-\eqref{NDYN_source} } showing precursors and wakes from a moving source with fixed charge. Here $v_{d} = 1.15$, $\phi_{d0} = 2.96$, $x_0 = 10$, $w = 0.50$ and $\sigma=0$. }
     \label{Fig_5}
 \end{figure}

In this case, we begin by first looking at the temporal behavior of the floating potential $\phi_d(t)$ as obtained by solving equations~(\eqref{NCE_eqn1} - \eqref{NPE_eqn3}) along with Eq.~\eqref{NCHGVd_eqn4}, with $v_d = 1.15$). The necessary condition to form the precursors, is that the source velocity must exceed the linear ion–acoustic wave phase speed $c_s$ (i.e. the Mach number is greater than unity).
In this context, $v_d=1.15$ corresponds to a weakly supersonic regime and is therefore well suited for clearly capturing the formation of precursor solitons. A lower threshold exists at $v_d \leq c_s$, where the system supports only wake-fields or linear wave excitations. Conversely, at much higher source velocities, the dynamics tend to become strongly nonlinear and may evolve into shock-dominated structures or pinned soliton formation~\cite{Garima_PRE_2021, Sanat_POP_2016A, Sanat_POP_2016B, Debkumar_POP_2022}. Consequently, precursor solitons are expected to arise only within a finite range of weakly to moderately supersonic source velocities, consistent with earlier studies~\cite{DesJardin_PRE_2025}.
\paragraph*{}
The present model takes into account the dynamic charging of the source. Figure~\ref{Fig_3} shows the temporal evolution of the floating potential on the moving debris source.
As in the case of the stationary source, the surface potential is seen to quickly acquire an equilibrium value close to $\phi_{d0} = -2.96$ in agreement with the approximate value derived in the previous section with an e-folding time of the order of $1.44$. The inset in Fig.~\ref{Fig_3} is a zoomed view of the early time dynamics of the charging process. As in the case of the stationary source, we also notice that $\phi_d(t)$ has small fluctuations in time around the equilibrium value $\phi_{d0}$ induced by the ambient plasma density and potential fluctuations. These low-frequency oscillations are again associated with ion-acoustic waves. 
\paragraph*{}
Next, we look at the excitation of precursors and wakes by the moving source that is dynamically charged by the plasma. Figure~\ref{Fig_4} shows the space-time evolution of the perturbed density driven by the moving source, which is being dynamically charged by the ambient plasma, that takes the form of precursor solitons ahead of the source and wakes behind the source. Similarly, Fig.~\ref{Fig_5} depicts a similar scenario when the floating potential on the source is not allowed to fluctuate but is held at the constant equilibrium value of $\phi_d(t) = \phi_{d0} = -2.96$. 
As can be seen from Figs.~\ref{Fig_4}  and \ref{Fig_5}, there is no significant difference in the excitation or dynamical evolution of the excited nonlinear structures, indicating that the charging dynamics does not have any significant impact on the generation of precursor solitons. The basic reason for this is the disparate time scales of the charging process and the soliton generation dynamics. As the early time simulations show, the charging process happens rapidly on the ion time scale whereas as we see from Figs.~~\ref{Fig_4} and \ref{Fig_5}, the solitons are created after several ion plasma periods. The charge and potential on the source remain fairly constant after the equilibration period with small temporal fluctuations which do not influence the generation of the solitons that occur on the longer ion-acoustic time scale. 
\paragraph*{}
It should be noted that while our present charging calculations are based on the OML theory~\cite{Goree_PSST_1994, Lai_IEEE_2012, Anderson_JGRSP_2012, Lhotka_POP_2020, Belanger_PRAB_2022}, in the space environment, depending on the region (altitude) the charging of space debris can be influenced not only by plasma currents but also by photoelectric emission~\cite{Sickafoose_PRL_2000, Munoz_PRL_2024, Taylor_POP_2024} and secondary electron emission~\cite{Richter_PRB_2006, Stefan_PRE_2006}. However, in the Low Earth Orbit (LEO) region that has the highest concentration of space debris, and is the principal focus of our work,  photoemission and secondary emission currents are typically not important in determining the charge state of debris objects, since they are much smaller than the particle fluxes of the ambient plasma. Photoemission contributions become important at higher altitudes e.g. in the Geosynchronous Earth Orbit (GEO) region. Being a more rapid phenomenon it does not alter the overall charging time and hence does not impact the excitation of precursor solitons. Secondary emission processes induced by energetic electron beams are important in the GEO and magnetospheric regions and can extend the charging time beyond the OML current charging time. However, the extended charging time still remains significantly shorter than the ion acoustic time scale and hence does not change the conclusions of our present calculations. 
Apart from charge fluctuation induced effects, precursor formation and propagation properties can also be influenced by other dissipative mechanisms such as ion Landau damping and collisional processes~\cite{Wong_PR_1964}. The ion Landau damping, a kinetic phenomenon, is beyond the scope of our present fluid model.  It can become important when the ion temperature is close to the electron temperature. Possible mechanisms that can overcome such damping include the existence of velocity shear in the plasma flow surrounding the moving charged object, as discussed in Ganguli {\it et al.}~\cite{Guru_POP_2025}. Likewise ion neutral collisions can cause damping of the linear ion acoustic fluctuations thereby delaying their buildup into larger amplitudes and the concomitant formation of the precursor solitons. They can further cause a damping of the soliton amplitude as it detaches itself from the driving source and propagates into the plasma.
\subsubsection{A numerical test of the Galilean invariance of the soliton generation process}
\label{PlasmaFlow}
In laboratory experiments on precursor solitons {\it e.g.} in dusty plasmas \cite{Surabhi_PRE_2016, Garima_POP_2019,Krishan_POP_2024}, it has been traditional to keep the charged source stationary and make the plasma flow towards it since the phenomenon is Galilean invariant. Our simulations in the next section will exploit this invariance. However, to verify that there are no numerical differences in the two approaches we have repeated the simulations of the previous section by making the plasma flow with $v_{p_f} = v_d$ towards a stationary source, where $v_d$ is the value of the source velocity used in the previous simulations. All other parameters have been kept the same.
The model system is then governed by the following set of coupled normalized evolution equations.
\begin{equation}\label{PF_NCE_eqn1}
\frac{\partial n}{\partial t} 
+\
\frac{\partial (n u)}{\partial x}
= v_{p_f} \frac{\partial n}{\partial x} 
\end{equation}
\begin{equation}\label{PF_NME_eqn2}
   \frac{\partial u}{\partial t} 
   +\ 
   u \frac{\partial u}{\partial x} 
   =\ 
 v_{p_f} \frac{\partial u}{\partial x} 
 -\frac{\partial \phi}{\partial x}  - \sigma \frac{\partial ln(n)}{\partial x}
\end{equation}
\begin{equation}\label{PF_NPE_eqn3}
    \frac{\partial^2 \phi}{\partial x^2}  
    =  e^{\phi} - n + S(x)
\end{equation}
The normalized form of the stationary source (\textit{i.e.,} $v_d = 0$) charged source density is given by
\begin{equation}
\label{PF_NDYN_source}
   S(x) = \frac{\phi_{d0}}{ \sqrt{\pi}} \exp\left[ -\left( \frac{ x-x_0}{w} \right)^2 \right]
\end{equation}
 \begin{figure}[h!]
     \includegraphics[width= \columnwidth]{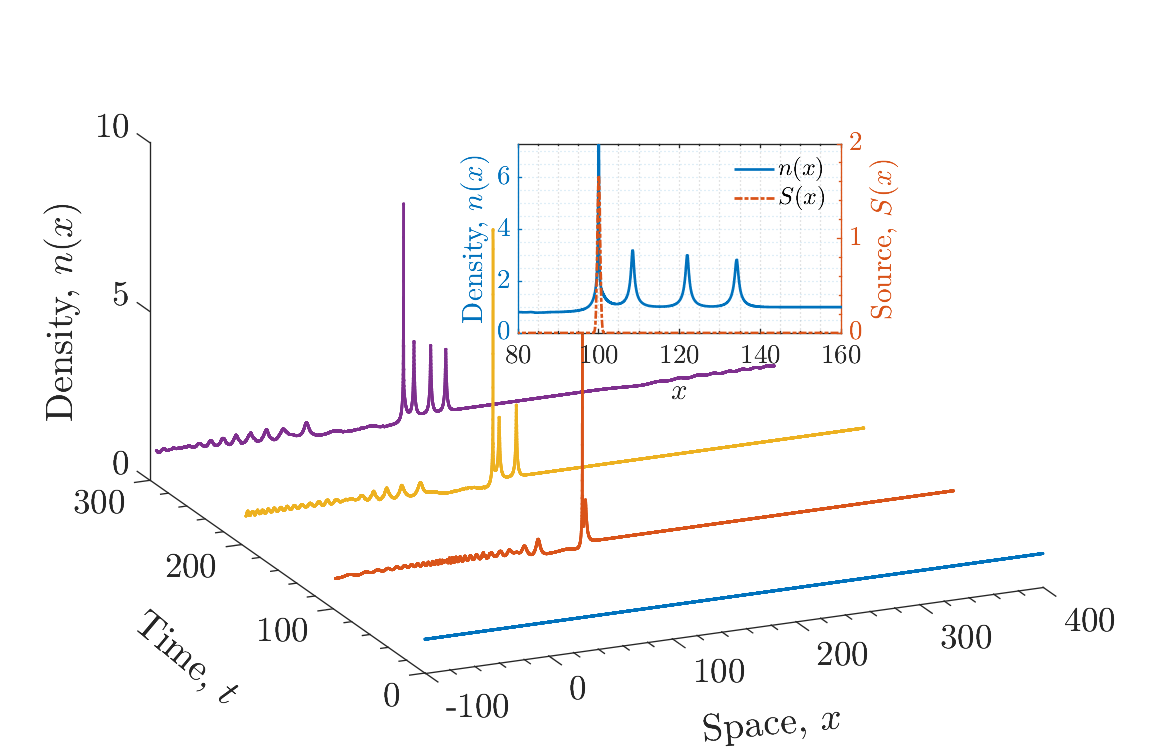}
     \caption{Spatio-temporal evolution of density using \eqref{PF_NCE_eqn1},~\eqref{PF_NME_eqn2},~\eqref{PF_NPE_eqn3}, and \eqref{PF_NDYN_source} showing precursors and wakes from a stationary source with fixed charge kept in a flowing plasma having flow velocity $v_{p_f} = 1.15$. Here $\phi_{d0} = 2.96$, $x_0 = 100$, $w = 0.50$ and $\sigma=0$.}
     \label{Fig_6}
 \end{figure}
The results are shown in Fig.~\ref{Fig_6}, and as can be seen the generated nonlinear coherent  structures (precursor solitons) and dispersive structures (wakes) are identical to the case where the Gaussian source was moving relative to the plasma {\it c.f.} Fig.~\ref{Fig_5}.
\begin{figure}
    \centering
    \includegraphics[width=1\columnwidth]{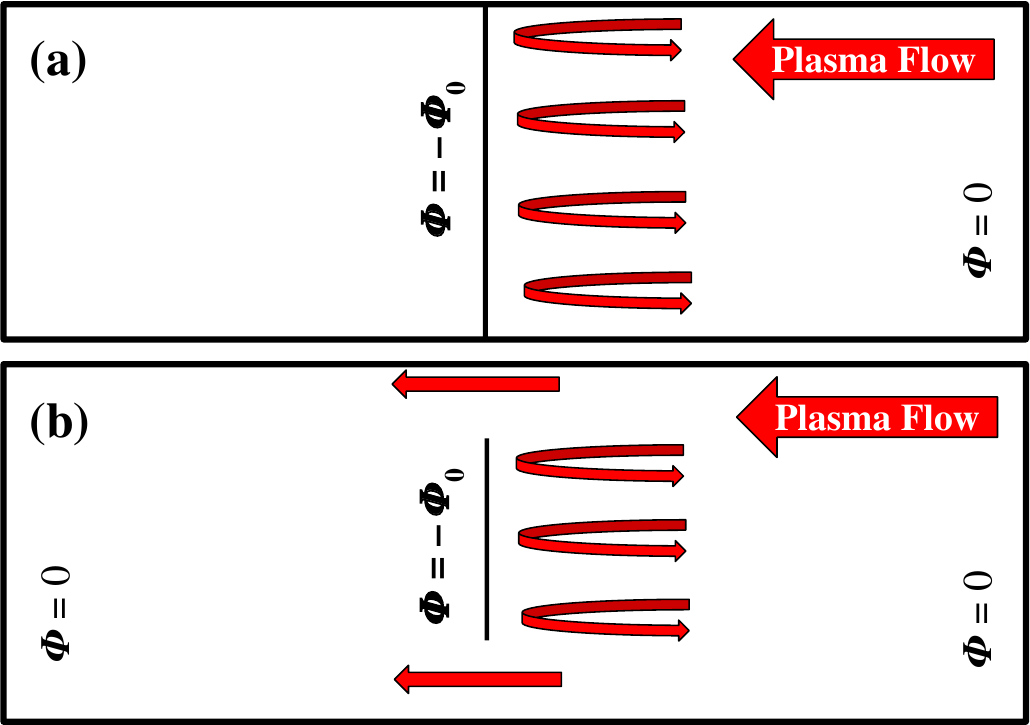}
    \caption{The cartoon representation of an infinite object (a) and a finite object (b) immersed in a plasma. Both are maintained at a fixed potential $\phi = - \phi_0$. The plasma is not able to penetrate the solid object. The  infinite object disconnects the plasma regions in the front and rear of the obstacle. The connectivity is restored for a finite object by allowing the plasma to flow over and below the object. 
    }
    \label{Fig_7}
\end{figure}
\section{Effect of an impermeable debris surface ON SOLITON FORMATION}
\label{Solid_Wall}
One of the characteristic features of a Gaussian source model, widely used in previous fluid and PIC simulations, is that it is transparent to the plasma, meaning that the plasma can flow through it. This is not physical, as in reality, the solid debris object will be impenetrable, and the plasma will have to flow around it. Would such a scenario influence the generation of precursors? Lira~{\it et al.}~\cite{Lira_JSR_2024} addressed this question in their PIC simulations by taking a one-dimensional source in the form of an infinite plane source. Instead of making the source (wall) move they made the plasma to flow towards the wall since the physical mechanism of soliton generation is Galilean invariant, as discussed in the previous  subsection. They found that when this plane source was subject to dynamical charging by the plasma, no precursor solitons were generated. Instead, a fluctuating sheath was seen to develop at the surface. In the next subsection we carry out a test of their PIC results in our fluid model.
\subsection{Sheath formation at an impermeable surface}
As in the PIC simulation we consider an infinite wall placed at the center ($x=0$) of the simulation box that is of extent $(-L,L)$ in the $x$ dimension (see Fig.~\ref{Fig_7}(a)). The partition that disconnects the right and left regions of the box represents the impermeable wall with the plasma reflecting off it. We consider only the $x:\{0, L\}$ half space since the infinite solid wall does not allow communication between the left and right halves of the physical space. Notice that the physical presence of the source is now represented through the boundary conditions at $x=0$.
\begin{figure}[h!]
    \centering
    \includegraphics[width=\columnwidth]{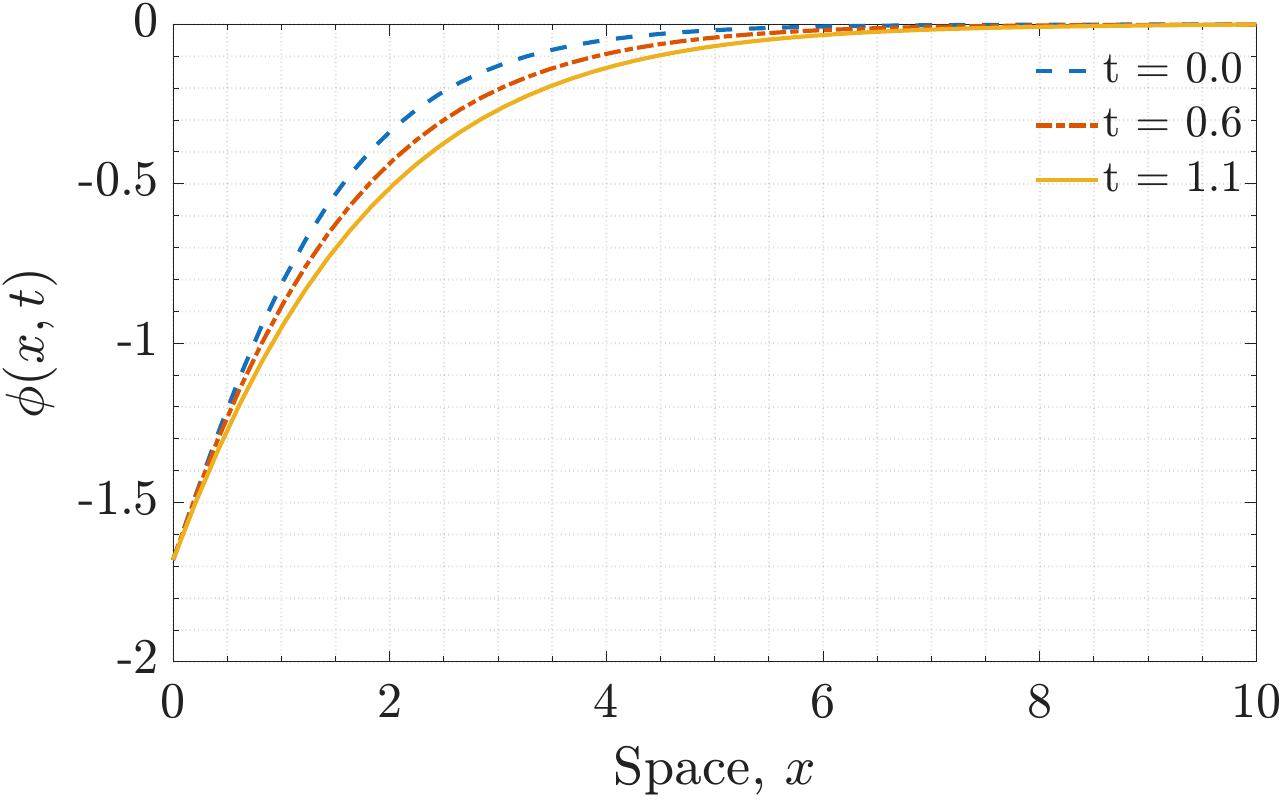}
    \caption{Sheath formation at the infinite wall maintained at potential $\phi_0 = -1.68$ with a plasma flowing towards it with a flow velocity $v_{p_f}=1.15$.}
    \label{Fig_8}
\end{figure}
\begin{figure}[ht!]
    \centering
    \includegraphics[width=\columnwidth]{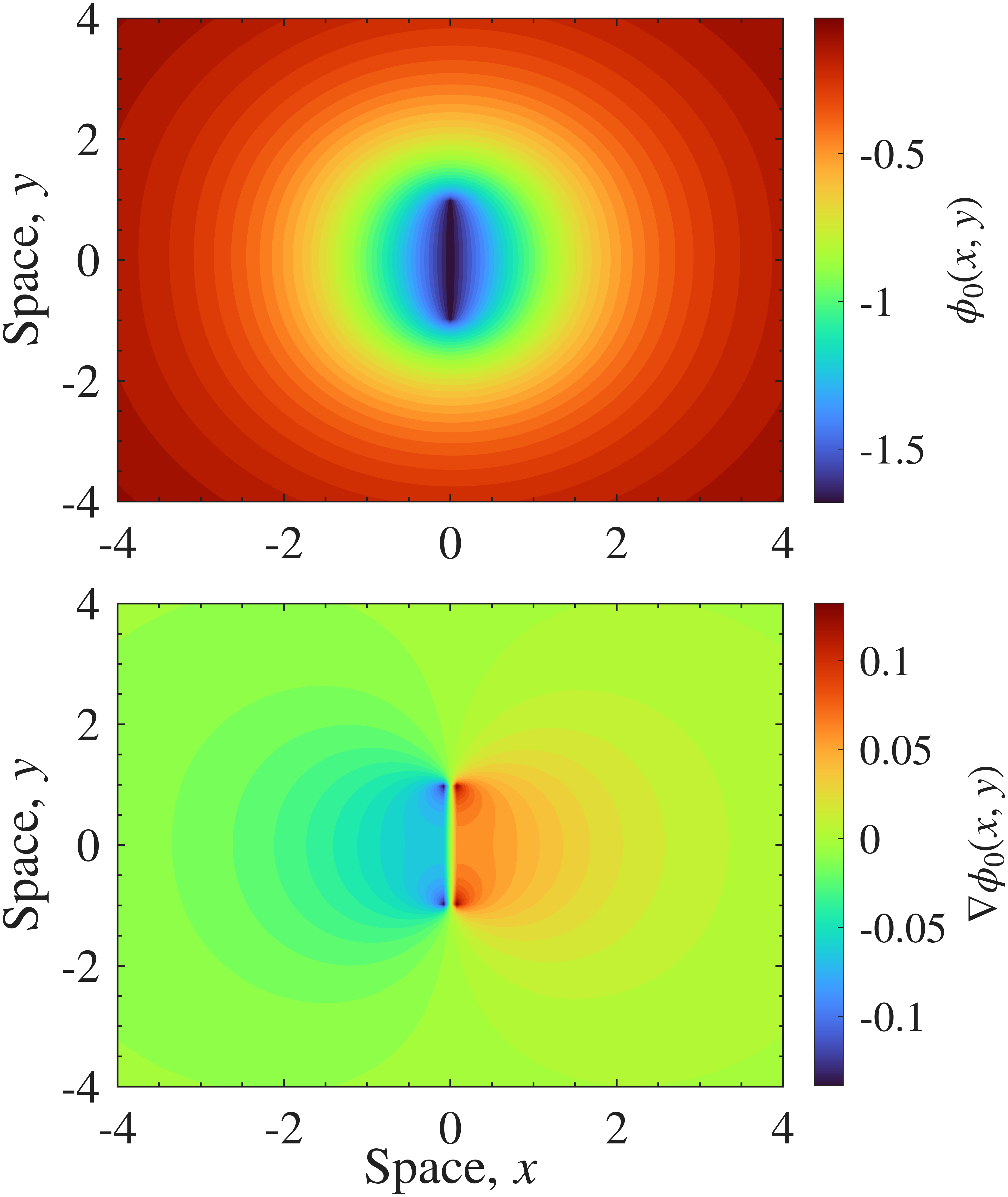}
    \caption{
    Magnified contours of the the potential field and the electric field around an impermeable finite line object maintained at a potential $\phi_0 = -1.68$. 
    }
    \label{Fig_9}
\end{figure}
\paragraph*{}
The following boundary conditions are applied.
$n(0,t) = 0 $;
$u(0,t) = 0 $;
$\phi(0,t)=\phi_0 = -1.68$;
$n(L,t)= 1$;
$u(L,t) = -v_{p_f}$;
$\phi(L,t) = 0 $.
Physically, this represents a plasma flowing with a velocity $v_{p_f}$ from the right towards a solid wall on the left at $x=0$,  that is kept at a potential $\phi_0=-1.68$, where $\phi_0$ represents the equilibrium floating potential of the infinite plane-shaped object in the plasma.
\paragraph*{}
We then solve the coupled systems of Eqs.~\eqref{PF_NCE_eqn1},~\eqref{PF_NME_eqn2}, and \eqref{PF_NPE_eqn3} simultaneously, by excluding the temperature effect $(\sigma=0)$ from the momentum equation and the charged source object from  the Poisson's equation. Following the findings of the Section~\ref{Charging_model}, we ignore for the present the dynamical charging of the wall and consider only the equilibrium value of the surface potential. 
A numerical solution of the above equations is displayed in Fig.~\ref{Fig_8}.
It shows the formation of a sheath at $x=0$ and no precursor solitons - a result akin to that of Lira~{\it et al.}~\cite{Lira_JSR_2024}. However, the basic reason for the absence of precursors is not the dynamical charging of the source but the absence of communication between the plasma in the front and rear of the source. For a Gaussian source, the precursor solitons were formed by pulling in mass from the rear of the source, which created a density depression behind the source. The infinite plane prevents such a transfer of mass and hence the absence of solitons. This can be remedied by taking a finite size source that would allow the plasma to flow around the object and connect the front and rear regions. To test this idea, we study, in the next section, the flow of a plasma on to a thin finite line object that does not allow the plasma to flow through it but allows it to flow over and below it. The problem is now two dimensional and also requires suitable modifications in the boundary conditions and model equations.
 \begin{figure*}[ht!]
     \centering
     \includegraphics[width=\textwidth]{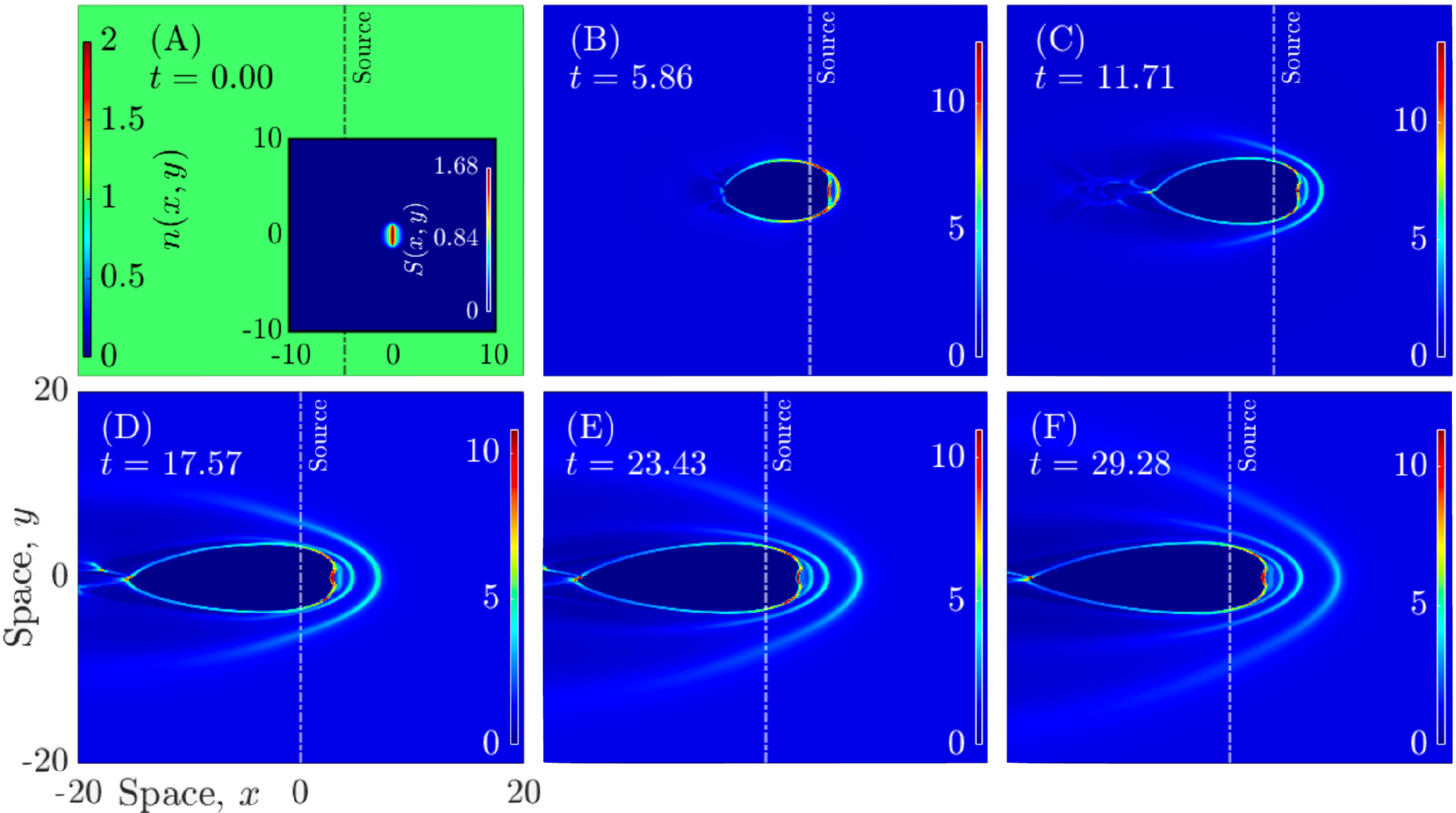}
     \caption{Density contours showing the generation of 2D electrostatic precursor solitons upstream of an impermeable finite object that is biased at a potential of $\phi_0 = -1.68$ with an oncoming flowing plasma with a velocity $v_{p_f}=1.15$. Subplot (A) shows the initial equilibrium state at $t = 0.0$, with a uniform plasma density $n(x,y) = 1$. The inset illustrates a localized, impermeable, finite-sized charged object held at a fixed potential and immersed in a supersonically flowing plasma. Subplots (B--F) present the temporal evolution of the system over $t = 5.86$--$29.28$, where ambient plasma perturbations progressively develop and ultimately give rise to precursor solitons excited by the presence of the finite charged object.
     }
     \label{Fig_10}
 \end{figure*}
\subsection{Model equations with a finite impermeable biased source}
\label{model_equations}
To simulate a finite impermeable source, we reconfigure the infinite wall, discussed in the previous subsection, by removing portions of the wall such as to leave a finite section in the center (see Fig.~\ref{Fig_7}(b)) which is biased at a fixed potential $\phi_0(0,0)$. This configuration allows the plasma to flow from the front to the rear through the gaps above and below the source. As in the previous subsection we allow the plasma to flow towards the source from the right to the left. By going to the frame of the flow we can obtain the following set of two dimensional normalized fluid equations:
\begin{widetext}
\begin{equation}\label{2D_NCE_eqn1}
        \frac{\partial n}{\partial t} + u_x \frac{\partial n}{\partial x} + u_y \frac{\partial n}{\partial y} 
        = 
        v_{p_{f}} \frac{\partial n}{\partial x} 
\end{equation}
\begin{equation}\label{2D_NMEx_eqn2}
       \frac{\partial u_x }{\partial t}     
     + u_x \frac{\partial u_x}{\partial x}
     + u_y \frac{\partial u_x}{\partial y}
     \\
     = 
     - \frac{\partial \phi}{\partial x}
     + v_{p_f} \frac{\partial u_x}{\partial x}  
     - \frac{\partial \phi_0}{\partial x}
\end{equation}
\begin{equation}\label{2D_NMEy_eqn2}
    \frac{\partial u_y }{\partial t} 
     + u_x \frac{\partial u_y}{\partial x}
     + u_y  \frac{\partial u_y }{\partial y} 
     = - \frac{\partial \phi }{\partial y}
     + v_{p_f} \frac{\partial u_y}{\partial x}
     - \frac{\partial \phi_0}{\partial y}
\end{equation}
\begin{equation}\label{2D_NPE_eqn3}
        \frac{\partial^2 \phi}{\partial x^2} + \frac{\partial^2 \phi}{\partial y^2}
        = \exp\left\{ \phi + \phi_0 \right\}  - n 
\end{equation}
\end{widetext}
where Eq.~\eqref{2D_NCE_eqn1} is the ion continuity equation, Eq.~\eqref{2D_NMEx_eqn2} and Eq.~\eqref{2D_NMEy_eqn2} are the $x$ and $y$ components of the ion momentum equation, and Eq.~\eqref{2D_NPE_eqn3} is the Poisson equation. The 2D field variables $n$,~$u_x$,~$u_y$, and $\phi$ represent the ion density, the $x$ and $y$ components of the ion velocity, and the electrostatic potential, respectively. $\phi_0(x,y)$ is the stationary potential arising from the biased line object. These dimensionless equations are obtained from the standard cold fluid model with the same approximations discussed for the 1D case in the Section~\ref{Charging_model} and follow the same normalization. 
\paragraph*{}
Note that instead of modeling the debris as an object with a fixed charge density, we have now recast the problem in terms of an object that is biased at a fixed potential, $\phi_0(0,0)$, representing the surface potential of the 2D charged debris.In the absence of a plasma such an object creates an electrostatic potential $\phi_0(x,y)$ around it that gives rise to an electric field $E_0(x,y) = -\nabla \phi_0(x,y)$. 
Numerical evaluations of the contours of $\phi_0(x,y)$ and $E_0(x,y)$ are shown in Fig.~\ref{Fig_9} (a), and Fig.~\ref{Fig_9} (b), respectively, for the configuration shown in Fig.~\ref{Fig_7}~(b).
This electric field provides an additional force on the ions that is now incorporated in the momentum equations \eqref{2D_NMEx_eqn2} and \eqref{2D_NMEy_eqn2}. 
It also influences the electrons which now have a modified Boltzmann distribution of the form $\exp\{\phi(x,y) +\phi_0(x,y)\}$, where $\phi(x,y)$ is the self-consistent plasma potential obtained by solving the Poisson equation \eqref{2D_NPE_eqn3} along with the dynamical equations \eqref{2D_NCE_eqn1}-\eqref{2D_NMEy_eqn2}. 
To ensure impermeability of the source the $x$ component of the velocity is made to reverse directions when it encounters the impermeable source (see Fig.~\ref{Fig_7}(b)). 
\subsection{Precursor Solitons using 2D Debris Simulation}
The above set of 2D model equations for nonlinear ion-acoustic excitations are solved in the simulation domain of dimensions $(-100\leq L_x \leq +100)$ in the $x$ direction and $(-20\leq L_y \leq +20)$ in the $y$ direction. The line source, placed at $x=0$ extends from $y=-1$ to $y=1$ and is biased with $\phi=\phi_0$. The plasma dynamical equations (\eqref{2D_NCE_eqn1} - \eqref{2D_NMEy_eqn2}) are solved using the FCT algorithm~\cite{Boris_FCT_1993} and the 2D Poisson equation is explicitly solved using the relaxation method~\cite{Sanat_POP_2016A}.
The following boundary conditions are used for the finite-sized (thin line) charged-debris: \\
$\phi(x=0,y \in \{-1,+1\},t) = \phi_0$;
$n(x=0,y \in \{-1,+1\},t)    = 1 $; 
\\
$u_x(x=0,y \in \{-1,+1\},t)  = - u_x(x=0, y \in \{-1,+1\} ,t)$.
\\
$\phi(x \neq 0,y \notin \{-1,+1\},t) = 0$;
$n(x \neq 0,y \notin \{-1,+1\},t)   = 1$;
$ u_x(x \neq 0,y \notin \{-1,+1\},t) = 0$; 
$ u_y(x \neq 0,y \notin \{-1,+1\},t) = 0$.
We next choose the same values of the plasma flow velocity and floating potential as were used for the infinite wall problem, namely, $v_{p_f} = 1.15$ and $\phi_0 = - 1.68$. 
\paragraph*{}
The findings of the 2D simulation of a finite-sized debris in a flowing plasma are shown in Fig.~\ref{Fig_10}. In contrast to the sheath structure seen in Fig.~\ref{Fig_8}, one observes the formation of distinct coherent structures of density perturbations traveling in the upstream direction of the impermeable object. 
The snapshot (A) in Fig.~\ref{Fig_10} at $t=0.0$, corresponds to the initial equilibrium density $n(x,y)=1$. The associated inset shows the charge density of the source (an impermeable finite object at a potential $\phi_0 = -1.68$). As the plasma flows supersonically over the finite-sized charged debris, perturbations develop, gradually becoming nonlinear and leading to the formation of precursor solitons, as illustrated in the later time snapshots (B-F) of Fig.~\ref{Fig_10}. 
These crescent shaped structures captured in snapshots (B-F) of Fig.~\ref{Fig_10}, are precursor solitons that are similar in character to those observed in earlier fluid and PIC simulations using a Gaussian charged source~\cite{Vikram_PRE_2023, Ajaz_PRE_2025}. As in those previous simulations one also observes the formation of a growing density depletion region behind the source and the occurrence  of trailing wake structures beyond that. Our present findings therefore demonstrate that the impermeability of the debris surface does not inhibit the formation of precursor solitons as long as the connectivity between the plasma regions in the front and rear of the object is maintained. It should also be noted that, as in previous simulations with a Gaussian source, the size of the depressed region at the rear of the object keeps growing as more and more solitons are created in the front as a consequence of mass conservation and is indicative of a close link between the dynamics of the two regions.
\section{Conclusions}
\label{conclusions}
To summarize, in this paper we have investigated two important physical effects that could potentially impact the formation of precursor solitons by a charged debris object, namely, the dynamical charging of the debris and the impermeable nature of its plasma facing surface. The dynamical charging effect has been studied through numerical solutions of an improved fluid model in which the time evolution of the charge on the debris has been accounted for in a self-consistent manner. A detailed comparison with earlier simulation results carried out for a source with a time independent charge shows that the charging process does not have any significant impact on the formation and evolution of precursor solitons. The physical reason for this is the wide disparity between the charging time and the formation or evolution time of the precursors. The charging time is primarily on the ion time scale while the solitons evolve on a longer ion-acoustic time scale. Our simulations reveal that the fluctuations of the floating potential developing on the source die away rapidly and the source acquires an equilibrium potential on an e-folding time scale that is consistent with the ion dynamics. The equilibrium floating potential then has small fluctuations representing linear ion-acoustic waves that are subsequently amplified by the moving source to create precursor solitons.
\paragraph*{}
To investigate the impact of an impermeable plasma-facing surface of the debris object, we carry out two simulation studies for identical values of the debris floating potential and the flow velocity of the plasma. In one case we use an infinite plane wall for the debris that essentially cuts-off the interaction between the front and rear region of the ambient plasma. In this situation one observes the formation of a plasma sheath at the wall surface as has been observed earlier in PIC simulations~\cite{Lira_JSR_2024}. However, when the simulation is repeated with a finite impermeable (line) source that permits the plasma to flow above and below it, one observes the formation of precursor solitons traveling in the upstream direction. These two dimensional crescent shaped solitons are akin to those observed in earlier fluid and PIC simulations using a simple Gaussian form for the debris source~\cite{Ajaz_PRE_2025, Vikram_PRE_2023}. One also observes the formation of a density depressed region and trailing wakes behind the source. 
Our findings demonstrate that the impermeable nature of the debris surface does not inhibit the formation of precursor solitons and thereby extends the practical applicability of the previous simpler models using a Gaussian form of the debris object. The present results also reveal the importance of having connectivity between the front and rear plasma regions surrounding the debris implying a deeper physical connection between the trailing wakes and the fore-wakes created by the moving object in the plasma. Our present results thus deepen our understanding of the formation and propagation characteristics of precursor solitons as observed in various past simulation~\cite{Sanat_POP_2016A, Atul_NJP_2020, Vikram_PRE_2023, Ajaz_PRE_2025} and controlled laboratory experiments~\cite{Surabhi_PRE_2016,Krishan_POP_2024, Garima_POP_2019}. These findings should prove useful in the current concerted efforts to unequivocally track their signatures in the space environment under the SINTRA program of the IARPA initiative \cite{Truitt_AMOS_2023, Bernhardt_POP_2023}.
\begin{acknowledgements}
This work has been supported by the Intelligence Advanced Research Projects Activity (IARPA) under its Space debris Identification and Tracking (SINTRA) program and also by the Air Force Office of Scientific Research under Grant No. FA9550-23-1-0003. A.S. is thankful to the Indian National Science Academy (INSA) for the INSA Honorary Scientist position. 
\end{acknowledgements}
\bibliography{Ref_DebrisCharging}
\end{document}